\documentstyle[aaspptwo]{article}
\tighten
\begin{document}
\title{Temperature Anisotropies and Distortions Induced by Hot
Intracluster Gas on the Cosmic Microwave Background.}

\author{F. Atrio--Barandela\altaffilmark{1}} 

\affil{F{\'\i }sica Te\'orica. Universidad de Salamanca. 37008 Spain.\\
	email: atrio@astro.usal.es}

\author{J. P. M\"ucket}

\affil{Astrophysikalisches Institut Potsdam. D-14482 Potsdam.\\
 	email: jpmuecket@aip.de}

\begin{abstract}
The power spectrum of temperature anisotropies
induced by hot intracluster gas on the cosmic background
radiation is calculated.  For low multipoles it remains constant 
while at  multipoles above $l>2000$ it is exponentially damped.
The shape of the radiation power spectrum is almost independent of the
average intracluster gas density profile, gas evolution history 
or clusters core radii; but the amplitude depends strongly on those parameters.
Its exact value depends on the global properties of the cluster 
population and the evolution of the intracluster gas. The distortion 
on the Cosmic Microwave Background blackbody spectra varies in a 
similar manner. The ratio of the temperature anisotropy to the
mean Comptonization parameters is shown to be almost independent
of the parameters of the cluster model and, in first approximation,
depends only on the number density of clusters. 
An independent determination of the contribution of clusters
to the distortion of the blackbody spectrum and the temperature
fluctuations of the Cosmic Microwave Background would determine the
number density of clusters that contribute to the Sunyaev--Zel'dovich effect.
\end{abstract}


\section {Introduction.}

The Compton scattering of the Cosmic Microwave Background (CMB) 
photons by the hot gas present in clusters of galaxies,
was first described by Sunyaev \& Zel'dovich (1970).
The Sunyaev--Zel'dovich  effect -hereafter SZ-
has two components: thermal and kinematic. The first
is caused by the random thermal motions of the electrons
whose distribution is assumed to be isotropic (in
the cluster reference frame). The kinematic component is an
additional effect due to the peculiar velocity of the cluster
with respect to the CMB (see Rephaeli 1995 for a review).
Nowadays, several groups are searching or are setting up
telescopes to measure the SZ effect on single clusters
(Holzapfel et al. 1997, Jones 1996, Saunders 1996).
But besides the effect of individual clusters the entire cluster
population acts as a screen of scattering sites through 
which the background
photons must pass to reach the observer. The overall effect
is both to distort the black body spectrum and to induce temperature
anisotropies on the CMB.
The latter is an important component of the
full anisotropy on angular scales of several arcminutes
(Cole \& Kaiser 1989, Bartlett \& Silk 1994, Colafrancesco et al. 1994).
The anisotropy depends on the cluster richness, its evolution and on
the cosmological model.  Therefore,  detection of 
anisotropies and distortions will yield significant insight into
the cosmological evolution of clusters. 

The analysis of the COBE/FIRAS data
yielded an upper limit on the degree of Comptonization
of the CMB (determined by the parameter $\bar y$ to be defined below)
of $\bar y\le 1.5 \times 10^{-5}$ at 95\% 
confidence level (Mather et al. 1994).
The Planck satellite, scheduled to be launched in the next
decade, will not specifically look for distortions.
Those will be deduced from the data on temperature
anisotropies. At the present stage several
groups are studying methods to substract foregrounds
(Tegmark \& Efstathiou 1996, Hobson et al. 1998)
and are performing simulations to predict the sensitivity
level at which cosmological parameters can be measured.
In this respect, it is necessary to carry out 
a careful analysis of the power spectrum of the foregrounds
since this will allow to distinguish at each frequency
between the genuine CMB distortions and
the contribution of clusters and other foregrounds.
As remarked by Tegmark \& Efstathiou (1996) this improves
the technique of removing foregrounds simply by comparing
data at different frequencies.
 
In this paper we shall obtain 
the radiation power spectrum of CMB temperature anisotropies
induced by the entire population of clusters of galaxies, the mean distortion
and the ratio of the temperature anisotropy
measured by SuZie (Church et al. 1997) to the Comptonization
parameter.  We will show that distortions and
temperature anisotropies are related not only for a single
cluster but  also for the values averaged over all
clusters. We extend the work of Bartlett \& Silk (1994)
and Colafrancesco et al. (1997)
by computing the power spectrum and analyzing its dependence
with the parameters describing the cluster population and the
number density. Contrary to the  latter authors, we
do not make a detailed analysis of all
cosmological models. We shall restrict our study to the
standard Cold Dark Matter model.

The outline of the paper is as follows: in Sec. 2,
we rederive the power spectrum of the temperature anisotropies 
due to the contribution of clusters of galaxies. 
In Sec. 3, we integrate our equations assuming a number density of 
clusters given by the Press-Schechter formalism.
We give the shape of the power spectrum and its dependence
on properties of the average SZ cluster population.
In Sec. 4, we show that while the amplitude of the 
radiation power spectrum and the CMB distortion 
varies strongly with the main cluster parameters, the
ratio of the temperature fluctuation to the mean Comptonization
parameter is almost independent of the parameters of the cluster model.
In first approximation, this ratio is determined solely 
by the number density of clusters.
Finally, in Sec. 5, we present our main conclusions.

\section{Temperaure anisotropies induced by clusters of galaxies.\label{sec2}}

The passage of CMB photons
through a single cluster of galaxies distorts the radiation spectrum 
due to the inverse Compton scattering. The amplitude of the
distortion depends on frequency. It can be described more appropriately
in terms of the change in brightness on the CMB. The temperature difference
is given by
\begin{equation}
{\delta T\over T_o} = g(x)y_c, 
\label{dtt}
\end{equation}
where $x=h\nu/kT$ is the CMB frequency in dimensionless units
and $g(x) = (x{\rm coth}(x/2) - 4)$ gives the frequency dependence
of the effect.  The cluster Comptonization parameter is defined as
\begin{equation}
y_c = {\sigma_T k_B \over m_e c^2} \int T n dl, 
\label{yc}
\end{equation}
where $n$ and $T$ are the intracluster electron density and temperature,
$\sigma_T$ is the Thomson cross section,
$k_B$ is the Boltzmann constant, $m_e$ is the
electron mass and the integral
is performed along the line of sight through the cluster.

Clusters are known to be extended X--ray
sources with gas density profiles well fitted by
\begin{equation}
n_e(r) = n_c(1+(r/r_c)^2)^{-3\beta/2},
\label{nr}
\end{equation}
where $n_c$ is the central electron density and $r_c$ is the core radius
of the cluster. The observed values of $\beta$,
obtained from X-ray surface brightness profiles, range from 0.5 to 0.7 
(Jones \& Forman 1984, Markevitch et al. 1997). 
Throughout this paper we will adopt $\beta=2/3$
since it permits a more simplified treatment.

The distortion induced by the IC gas depends on the 
temperature profile along the line of sight through the cluster. 
While most authors assumed the IC gas to be isothermal,
in a recent study encompassing 26 nearby ASCA clusters  
Markevitch et al. (1997) 
have shown that the gas temperature within clusters decreases
slowly with radius. Note that the uncertainty about the
ASCA point spread function implies that errors on the
density profile of individual clusters are correlated. Therefore,
their data can not be used to estimate the average properties
of the IC gas. To simplify, we shall retain the
hypothesis of isothermality since the computation becomes simpler,
but keeping in mind that our analysis could be easily
modified to take into account any dependence of temperature with radius. 

If the virial radius of the cluster $r_v$ is expressed 
in terms of the core radius as $r_v=pr_c$,
eq.~[\ref{yc}] can be recast in the following manner:
\begin{equation}
y_c	= y_o\phi(\theta),
\label{ycf}
\end{equation}
with $y_o = [k_B\sigma_T/m_ec^2]r_c T n_c$,
$\theta$ the angular separation between the line
of sight and the center of the cluster and
\begin{equation}
\phi(\theta) = {2\over\sqrt{1+(\theta/\theta_c)^2}}
\tan^{-1}\left[\sqrt{{p^2 - (\theta/\theta_c)^2}
	\over{1+(\theta/\theta_c)^2}}\right] .
\label{phi}
\end{equation}
In this last expression, $\theta_c$ is the angle subtended
by the core radius of the cluster.
Eq.~[\ref{ycf}] gives the effect produced by a single 
cluster through a particular line of sight. Note that $p$, given
as the ratio between the virial and the core radius, is not a parameter but
a function of mass $M$ and redshift $z$.

\subsection{Mean Comptonization Parameter.}

The mean distortion on the CMB induced by 
all clusters is obtained adding the effect
of one cluster for all possible lines of sight:
\begin{equation}
\bar y = \int {dn\over dM}dM{dV\over dz}dz \kappa y_o \bar\phi .
\label{ycmean}
\end{equation}
In this expression, $dn/dM$ is the cluster number density
per unit of mass,
$\kappa$ gives the probability that a particular 
line of sight crosses a cluster. This probability is simply 
$\kappa = (p \theta_c)^2 /4$. Finally,
$\bar\phi$ is the averaged line of sight through
a cluster:
\begin{equation}
\bar\phi = {\int \phi \theta d\theta\over  \int \theta d\theta}
	 = 4p^{-2}(p-\tan^{-1}p) .
\label{phim}
\end{equation}
It is useful to introduce the following notation:
\begin{equation}
< \phi > \equiv {1\over 2}\int \theta d\theta \phi .
\label{average}
\end{equation}
This allows to write $\kappa \bar\phi = <\phi>$.
Now eq.[\ref{ycmean}] can be rewritten as follows:
\begin{equation}
\bar y = \int {dn\over dM}dM{dV\over dz}dz y_o <\phi> .
\label{ybar}
\end{equation}
Notice that $\bar y$ can now be understood as the 
average of $y_o <\phi>$ over the entire cluster population.
The superposed effect is determined
by the cosmological model, cluster abundance and redshift
evolution of the IC medium (Colafrancesco et al. 1997). 
 
\subsection{Radiation Power Spectrum.}

The IC gas does not only produce  distortions on the CMB spectrum.
It also induces temperature anisotropies.
The contribution of a single cluster 
to the temperature anisotropy on a wavenumber $\bf l$ is 
\begin{equation}
{\delta T\over T}({\bf l})= {1\over 2\pi}g(x)y_o\int d^2\theta
%\mbox{\boldmath$\theta$}
	\phi(\theta)e^{-i{\bf l}\mbox{\boldmath$\theta$}} .
\end{equation}

The power spectrum can be obtained
by adding in quadrature the contribution of all clusters.
Neglecting cluster spatial correlations,
we can write (Cole \& Kaiser 1988; cp. also Bartlett \& Silk 1994)
\begin{equation}
P(l)	= \int {dn\over dM}dM{dV\over dz}dz
	(g(x)y_o)^2|\tilde \phi(l)|^2,
	\label{pq}
\end{equation}
where $\tilde \phi(l)$ is the Fourier transform of the angular
profile of the cluster.
For scales much larger than the virial radius of a typical cluster, radiation 
temperature anisotropies originated by the contribution of
Poisson distributed clusters lead to $P(l)\simeq  const$.

For scales $l>\theta_c^{-1}$ the behaviour of $P(l)$
can be obtained analytically taking the limit $p$ going
to infinity (but keeping $p\theta_c$ finite). 
In this limit, the radiation power spectrum becomes exponentially
damped, that is, it decreases faster than $l^4$ as 
expected by Tegmark \& Efstathiou (1996).
In agreement with these authors, we shall demostrate numerically
that $l^2P(l)$ has a coherence scale, i.e. a maximun, around
$l\simeq 1000 - 2000$.  For reference, let us mention that the angular scale 
$l = 2000$ corresponds to a wavenumber $k \simeq 0.5h$Mpc$^{-1}$
or scalelength $\lambda \simeq 10h^{-1}$Mpc. 

From the power spectrum we can compute the
correlation function of temperature anisotropies. In the 
flat sky approximation, it is
given by (Atrio-Barandela, Gottl\"ober \& M\"ucket 1997)
\begin{equation}
C(\alpha)={1\over 2\pi}\int ldlW(l)P(l)J_o(l\alpha),
\label{correlation}
\end{equation}
where W(l) represents the window function of the experiment.
This expression can be rewritten in a more illuminating form.
Replacing the power spectrum (eq.~[\ref{pq}]),
and after some algebra, for an experiment
with infinite resolution and full sky coverage, we obtain:
\begin{equation}
C(0)= {1\over\pi} g(x)^2 \int {dn\over dM}dM{dV\over dz}dz
        y_o^2 < \phi^2 > ,
\label{correlation_analytic}
\end{equation}
where $\phi$ is given by eq.~[\ref{phi}] and the
average was defined in eq.~[\ref{average}]. This expression
states that the correlation 
function at the origin represents the average
of $y_o^2<\phi^2 >$ over all clusters. 
For a single cluster, Compton
parameter $y_c$ and temperature anisotropies 
are related (see eq.~[\ref{dtt}]). 
Comparison with eq.~[\ref{ybar}] suggest that a
similar relation holds 
between $\sqrt{C(0)}$ and $\bar y$. In Sec. 4 we shall
show that this is indeed the case. We shall particularize
this relation for the SuZie experiment (Church et al. 1997) 
and with the upper limits obtained from their observations
we shall set a constraint on $\bar y$ stronger than
the one derived from COBE/FIRAS.

\section{Cluster Model.}

Numerical estimates of the mean Comptonization parameter,
the power spectrum and the correlation function for a given
experiment require the cluster number density and average properties,
as a function of mass, redshift evolution of
the IC gas and cosmological model, to be specified.
The number density of clusters per unit of redshift is given
by the Press--Schechter formula (Press \& Schechter, 1974):
\begin{equation}
{dn\over dM} = \sqrt{2\over\pi}{\rho_b\over M}{\delta_v b\over\sigma^2}
        {d\sigma\over dM}\exp(-{\delta_v^2 b^2\over 2\sigma^2}),
\end{equation}
where $\rho_b$ is the background density at redshift $z$,
$M$ is the virial mass of the cluster, $\sigma$ is the rms of the linear density
fluctuation field at $z$, smoothed over the region containing $M$, $b$
is the bias factor and
$\delta_v$ is the linear density contrast of a perturbation that
virializes at $z$. The background density and the variance of the
density field scale with redshift: $\rho_b \sim (1+z)^3, \sigma\sim (1+z)^{-1}$
while the bias and $\delta_v$ are assumed to be constant. 

We shall restrict our analysis to a Cold Dark Matter model
with Hubble constant $h=0.5$ ($H_o = 100h$ km s$^{-1}$Mpc$^{-1}$)
and normalized to produce
a rms matter density perturbation at a $8h^{-1}$Mpc scale
of $\sigma_8 = 0.7$. Only for this particular model and
using the largest numerical simulation to date
Tozzi \& Governato (1997) checked that 
for clusters with masses  larger than $10^{14}$M$_\odot$,
the Press-Schechter formula
was in reasonable agreement with the number density of clusters 
up to redshift $z=1$ if $\delta_v b = 2.1$.  A slightly larger
value ($\delta_v b= 2.6$) was found by Colafrancesco \& Vittorio (1994)
by fitting the standard CDM model to the observed cluster
X-ray luminosity function. The results presented
in the next section were obtained assuming the former value 
at all redshifts. 

To compute the temperature anisotropies induced by hot
IC gas, we need to translate the properties of  a sample of clusters
at low redshifts into their equivalents at earlier
epochs. In what follows, the virial mass $M$ will be expressed
in units of $10^{15}$M$_\odot$.  For the spherical collapse model,
the virial radius scales as $r_v= r_{vo} M^{1/3}(1+z)^{-1}$
where $r_{vo}$ is the current
average virial radius of a $10^{15}$M$_\odot$ cluster.
We shall use the entropy-driven model of cluster evolution developed by
Bower (1997) to describe the core radius evolution as a function of redshift 
and mass. This model is applicable to an isothermal IC gas distribution.
If temperature is proportional to the velocity 
dispersion of the dark matter then
\begin{equation}
T  = T_g M^{2/3}(1+z) ,
\label{Temp}
\end{equation}
where $T_g$ is a normalization constant. It corresponds to the current
temperature of the IC gas of a cluster of $10^{15}$M$_\odot$.
In our numerical estimates we took $T_g = 10^8$K.
The core density evolves as
\begin{equation}
\rho_c \propto T^{3/2} (1 + z)^{-3\epsilon/2} ,
\label{cd}
\end{equation}
where $\epsilon$ parametrizes the rate of core entropy evolution.
Therefore, the central electron density 
scales as: $n_c = n_{co} (T/T_g)^{3/2} (1 + z)^{-3\epsilon/2}$.
From their study on the
Luminosity - Temperature relation on clusters at high redshift,
Mushotzky and Scharf (1997) found $\epsilon = 0\pm 0.9$. 

For a cluster with $\beta = 2/3$, assuming that $r_v \gg r_c$,
eq.~[\ref{cd}] leads to the following expression for the core radius:
\begin{equation}
r_c = r_{co} M^{-1/6} (1 + z)^{(-1+3\epsilon)/4} ,
\label{rc}
\end{equation} 
where $r_{co}$ is the average core radius today for a cluster
of $10^{15}$M$_\odot$. From this expression one obtains 
$p = p_o M^{1/2}(1+z)^{-{3\over 4}(1+\epsilon)}$. In our numerical estimates, we took
$r_{vo} = 1.3$h$^{-1}$Mpc. The core radius today is 
given by $r_{co}=r_{vo}/p_o$. In our analysis we have considered
the effect of different core radii, i.e.,
$p_o= 7,10,15$,  since, according to Makino, Sasaki \& Suto (1997),
the core radius of clusters is uncertain and could have been overestimated. 
Finally, we took the present gas density  of a $10^{15}$M$_\odot$ cluster
to be $n_{co} = 2\cdot 10^{-3}$cm$^{-3}$ (Peebles, 1993)
for a Hubble constant of $h=0.5$.

We would like to remark that all the 
relations introduced in this section are 
to be thought accurate in determining the
mean evolution of the cluster population as a whole. They should
not be considered applicable to individual clusters.

\section{Results and Discussion.}

We obtained the power spectrum of temperature anisotropies
induced by clusters of galaxies by numerically 
integrating eq.~[\ref{pq}]. 
The main limitation of our work is to determine
the scale of the less massive objects containing enough hot
gas to produce a measurable effect on the CMB. 
We shall ellaborate on this point further
below. Except otherwise specified, we assumed 
a lower limit $M_l=10^{14}$M$_\odot$ since above that mass scale
our cluster model is an adequate description of clusters
in the local Universe (Mushotzky \& Scharf, 1997)
and the Press--Schechter formula has been found
to be accurate (Tozzi \& Governato, 1997).  As an
upper limit we took $M_u=2\cdot 10^{15}$M$_\odot$.
 
\begin{figure*}[t]
\vspace{-1cm}
\centerline{\epsfxsize=16cm %\epsfysize=9cm
            \epsfbox{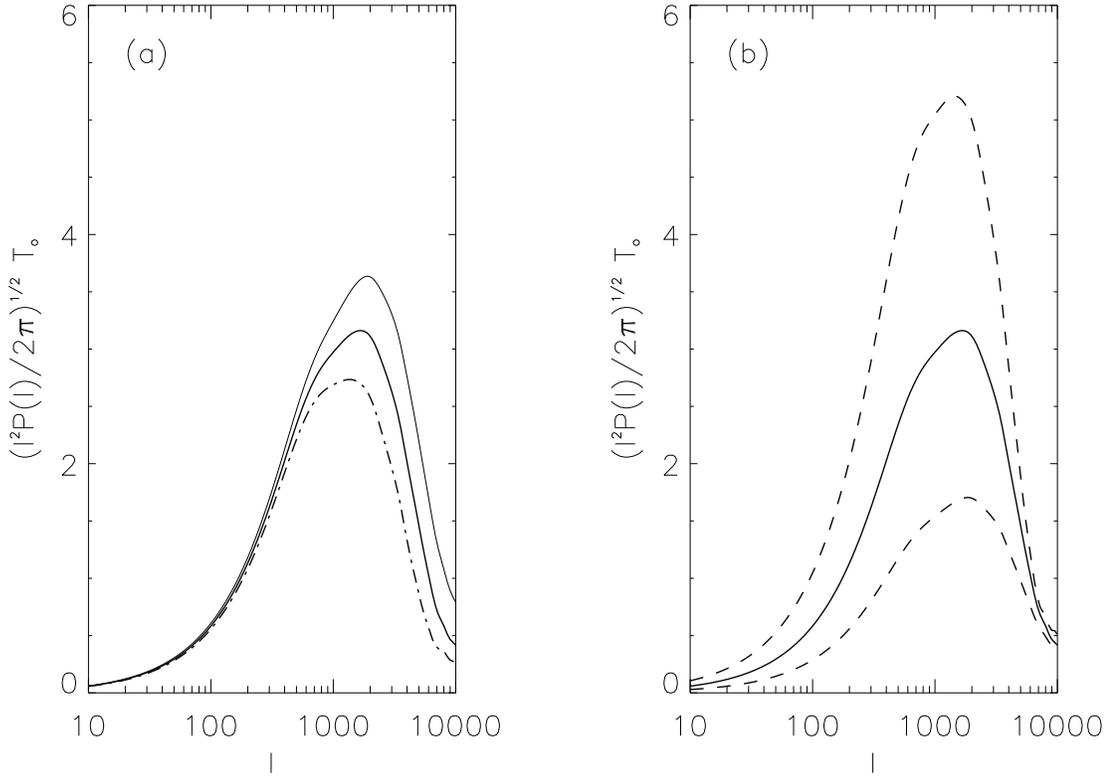}		 }
\vspace{-0.5cm}
\caption[FIG. 1]{Mean CMB temperature offset
induced by hot gas in clusters of galaxies expressed in units of $\mu$K.
In (a) plots correspond to different gas evolution histories
with identical virial to core radius ratio today ($p_o = 10$):
thin solid line $\epsilon = -1$, thick solid line $\epsilon = 0$ and
dot-dashed line $\epsilon = 1$. In (b) 
temperature fluctuations for different ratios 
of virial to core radius today with $\epsilon = 0$
are represented: upper dashed line
$p_o = 7$, thick solid line $p_o = 10$ and lower dashed line $p_o = 15$.
Integration of eq.[~\ref{pq}] was performed in
the mass range $10^{14}$M$_\odot < M < 2\cdot 10^{15}$M$_\odot$.
}
\label{fig1}
\end{figure*}

In Fig.~(\ref{fig1}) we plot the mean CMB temperature 
offset $T_o(l^2P(l)/ 2 \pi)^{1/2}$. The
CMB blackbody temperature $T_o$ is expressed in units of $\mu$K.
We took $g(x) = 1$ in all our plots, so the y-axis should
be multiplied by the correct value when comparing with 
a particular experiment.
One should notice that the mean temperature offset scales
as: $T_g r_{co}n_{co}g(x)$ where the values of all these
constants, given in the previous section,
correspond to a $10^{15}$M$_\odot$ cluster. In this way, the y-axis
on Fig.~(\ref{fig1}) can be easily rescaled to a different set of values.
All plots correspond to CDM ($\Omega_B = 0.05$, $H_o$ = 
50 km s$^{-1}$Mpc$^{-1}$) normalized $\sigma_8 = 0.7$.
In Fig.~(\ref{fig1}a) we consider the effect of different IC
gas evolution histories: $\epsilon=-1$ (thin solid line), $\epsilon=0$
(thick solid line) and $\epsilon=1$ (dot-dashed line). In all plots
we considered the ratio of virial to core radius to be $p_o = 10$ today.
In Fig.~(\ref{fig1}b) we show the effect of varying the core radius for a fixed
virial radius today, and no
IC gas evolution ($\epsilon=0$). We considered $p_o = 7,10,15$. Notice that
the amplitude is largest for $\epsilon = -1$ and
$p_o = 7$ since those models lead to larger gas fractions. As expected from
our discussion in Sec. 2, all power spectra
have a similar bell shape in all cases. The coherence
scales are around $l=1000-2000$. Differences in shape are 
less important than in amplitude. 
Also, notice that the dependence is stronger on the core radius
than on the IC gas evolution. If our cluster model
is correct, measuring the anisotropy
induced by clusters of galaxies an estimate of the
average core radius of the cluster population
could be obtained which will only depend marginally 
on the IC gas evolution. 

\begin{figure*}[t]
\vspace{-1cm}
\centerline{\epsfxsize=16cm %\epsfysize=9cm
            \epsfbox{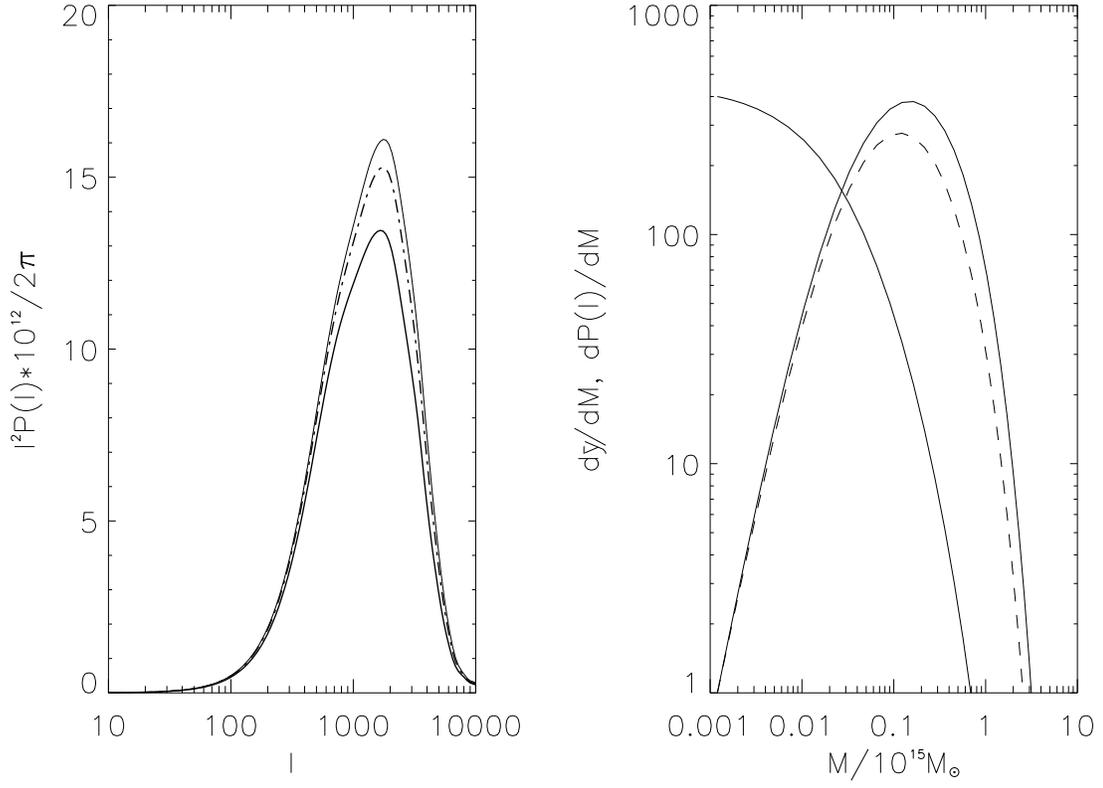}
		 }
\vspace{-0.5cm}
\caption[FIG. 2]{(a) Power spectrum of CMB temperature
anisotropies induced by hot IC gas.
The curves differ in their lower limit in
the mass integration of eq.~[\ref{pq}].
Solid line correspond to a lower limit
of $10^{13}$M$_\odot$, dot-dashed line to
$5\cdot 10^{13}$M$_\odot$ and thick solid line to $10^{14}$M$_\odot$.
The mass upper limit was the
same as those of Fig.~[\ref{fig1}]. (b) Left solid line: $d\bar y/dM$.
Dashed line $dP(l)/dM$ for $l=100$ and right solid line for $l=1000$.
The scale on the y-axis is arbitrary and the $d\bar y/dM$ 
curve has been rescaled to fit in the frame.
}
\label{fig2}
\end{figure*}

In Fig.~(\ref{fig2}a) we show the power spectrum of temperature
anisotropies for different
mass cut-off: $M_l = 10^{13}, 5\cdot 10^{13}, 10^{14}$M$_\odot$.
Notice that the amplitude of the power spectrum is
remarkably unsensitive to the change in the lower limit,
the reason being that the integrand of eq.~[\ref{pq}] is peaked
around $2\cdot 10^{14}$M$_\odot$ and falls there of, as indicated
by Fig.~(\ref{fig2}b). In this figure, the left solid line displays  
$d\bar y/dM$, the dashed line represents $dP(l)/dM$ for $l=100$ while
$l=1000$ is represented
by the right solid line.  As a result, the bulk of the contribution
to the radiation power spectrum comes from clusters of mass close to $10^{14}$M$_\odot$.
So, even though the entropy driven
model of clusters we used has not been checked for cluster
masses below $10^{14}$M$_\odot$, the CMB temperature fluctuations
shown in Fig.~(\ref{fig1}) are reasonable accurate.
This is not the case for the mean Comptonization parameter
$\bar y$ which depends strongly on the lower mass cut-off
as the contribution of clusters of small mass dominate the
integral, as indicated by Fig.~(\ref{fig2}b).

\begin{figure*}[t]
\vspace{-1cm}
\centerline{\epsfxsize=16cm %\epsfysize=9cm
            \epsfbox{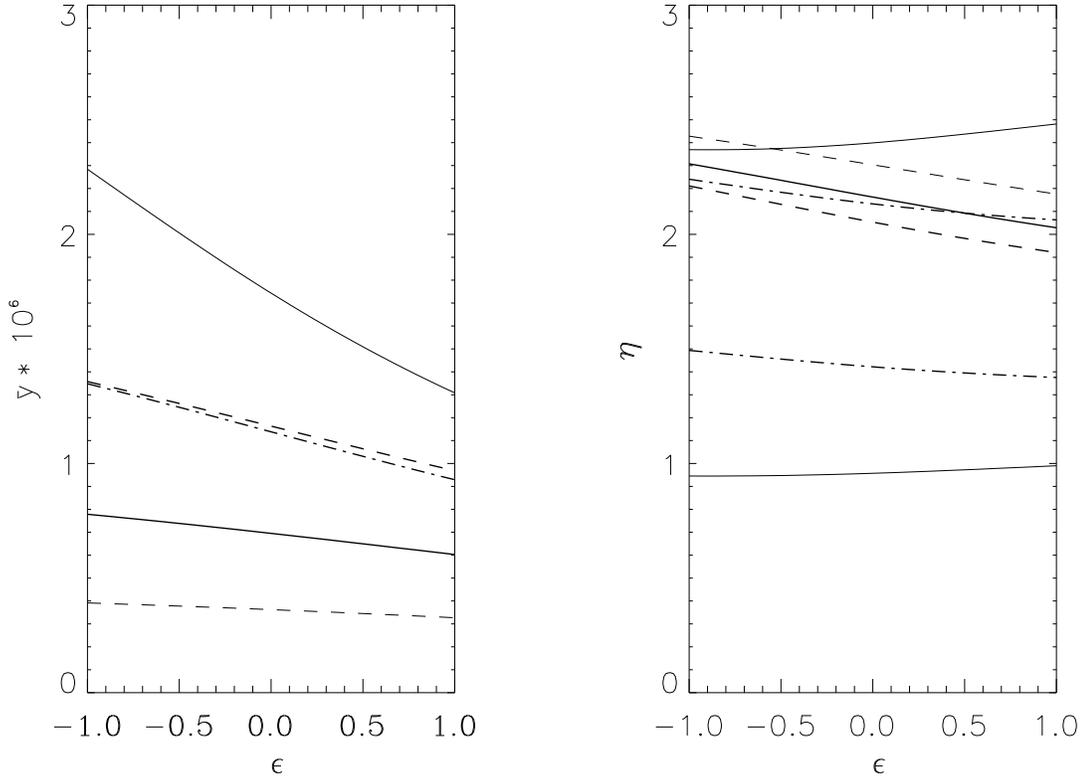}		 }
\vspace{-0.5cm}
\caption[FIG. 3]{
(a) Mean Comptonization parameter.
Thin solid, dot-dashed and thick solid
lines correspond to $p_o = 10$ and lower
limit of mass integration $10^{13}, 5\cdot 10^{13}$
and $10^{14}$M$_\odot$, respectively.
Upper and lower dashed curves correspond to
$p_o = 7, 15$ and $10^{14}$M$_\odot$ lower limit.
(b)Ratio of $\sqrt{C(0)}$ to the mean Comptonization
parameter. The two bottom curves correspond to
different lower limits on the mass integral
(thin solid line $10^{13}$M$_\odot$ and
dot-dashed line $5\cdot 10^{13}$M$_\odot$). In this two
cases the quantity $\eta\sqrt{\bar n/n_{cl}}$ was also plotted
(upper thin solid and dot-dashed lines).
We took
the number density of SZ clusters to be
$n_{cl} = 2.5\cdot 10^{-4}h^{3}$Mpc$^{-3}$.
The upper dashed, thick solid and lower dashed
lines correspond to $p_o = 15, 10, 7$, respectively.}
\label{fig3}
\end{figure*}

In Fig.~(\ref{fig3}a) we plot the mean Comptonization
parameter as a function of the 
IC gas evolution. We varied the lower mass limit of the
integral in eq.~[\ref{ycmean}].
From top to bottom, the thin solid line corresponds to 
a lower mass limit of $ M_l = 10^{13}$M$_\odot$, the dash-dotted
line to $5\cdot 10^{13}$M$_\odot$, and the thick solid
line to $10^{14}$M$_\odot$, all three with $p_o = 10$.
The upper and  lower dashed lines correspond to
$p_o = 7, 15$, respectively. In the latter two cases 
the lower mass limit was $10^{14}$M$_\odot$.

In analogy to the relation eq.~[\ref{dtt}] we introduce a 
new variable $\eta$  defined as
\begin{equation}
\eta \equiv {\sqrt{C(0)} \over \bar y }
\label{eta}
\end{equation}
In Fig.~(\ref{fig3}b) we plot the behaviour of $\eta$
for the most extreme parameters of the cluster model used. 
To calculate the correlation function we used SuZie window 
function. Again, we assumed $g(x) = 1$.
Lines correspond to the same models as in Fig.~(\ref{fig3} a).
The curves corresponding to different lower limits on the mass 
integral are represented twice. We have ploted
$\eta$ (lower thin solid and dot dashed lines) 
and $\eta\sqrt{\bar n/n_{cl}}$
(upper thin and dashed-dotted lines).
This figure demonstrates that while $C(0)$ and $\bar y$
depend strongly on the evolution of the IC gas and
the size of the core radius, $\eta$ varies by
less than 20\%. 
The largest dependence of $\eta$ comes
from varying the present number density of clusters. 
In the Press-Schechter theory,
decreasing the lower mass limit
increases $\bar n$, the number density of clusters 
above a given mass scale.
In our case, the lower mass limits $ M_l = 10^{13}, 5\cdot 10^{13}, 10^{14}$M$_\odot$
correspond to $1.5\cdot 10^{-3},
6\cdot 10^{-4}$ and $2.5\cdot 10^{-4}$ clusters per
$h^{-3}$Mpc$^3$, respectively.
If $C(0)$ and $\bar y$ instead of being normalized to
$\sigma_8$ were normalized to
a fixed number of clusters today, temperature correlation
and mean distortion would be rescaled as: 
$C(0)\cdot n_{cl}/\bar n$ and $\bar y\cdot n_{cl}/\bar n$,
where $n_{cl}$ is the number density of clusters that
produce a measurable effect on the CMB.
Then $\eta$ will scale as $\sqrt{\bar n/n_{cl}}$.
Going back to Fig.~(\ref{fig3}b)
the reader can check that 
this rescaling renders the solid thin and the dot-dashed
lines in the range of $\eta \simeq 2.0$.

Let us remark that the value of $\eta$ is different for different
experiments. For Suzie $\eta\simeq 2.0$ and for
an experiment with infinite angular resolution $\eta\simeq 2.5$
(see eq.~[\ref{correlation_analytic}]).
However,  a tight relation between
distortion and anisotropy exists, independently
of the experiment considered. This tight relation is
the consequence of the unsensitivity of
the power spectrum with the lower limit of the mass
integral. 

The upper limits on CMB temperature
anisotropies obtained by SuZie (Church et al. 1997)
on angular scales of $l\simeq 2000$ can be used to
estimate upper limits on the mean Comptonization
parameter that are stronger than those set by
COBE/FIRAS. Let us assume that that temperature
anisotropies on scales measured by SuZie are dominated
by clusters, a very conservative assumption for the argument that
follows.  Using the results
summarized in Fig.~(\ref{fig3}b), we can impose
an upper limit on $\bar y$ from SuZie 
upper limit (in absolute value) $\sqrt{C(0)} \le 2.1 \times 10^{-5}$
at the 95\% confidence level. 
At 142 GHz, SuZie operating frequency, 
$g(x) \simeq -1$. If we assume that only clusters
with masses larger than $10^{14}$M$_\odot$ contribute
to the CMB distortion, we have $\eta \ge 2.0$.
Therefore, $\bar y \le  1.0 \times 10^{-5}$,
a contraint stronger than the upper limit
obtained by Mather et al. (1994) at the same confidence level. 
Let us remark that the SuZie upper limits quoted above
were obtained assuming Gaussian statistics, while in
our analysis we assumed that clusters were Poisson distributed
on the sky. A firm upper bound can not be obtained
wihtout reanalyzing the SuZie data. 

The mean density of clusters can be estimated with 
independent measurements of distortion and temperature
anisotropy. If the contribution of clusters to the
temperature fluctuations of the CMB is measured by, say, SuZie
and the blackbody distortion is also determined, 
comparison with Fig (3b) would allow a direct estimate of
the number density of clusters with IC gas hot
enough to produce a significant contribution to the SZ effect. 
Comparison of the mean number density of SZ clusters with
that of X-ray clusters would lead to a better understanding of the
IC gas evolution history.

\section{Conclusions.}

In this paper we have shown that the shape of the power spectrum
of temperature anisotropies induced by the hot IC
gas is almost independent of model parameters. However, the amplitude
does depend on how the mean population of clusters is modeled.
As shown by Tegmark (1998) the shape of the
power spectrum is important for extracting 
the foreground contribution on experiments
measuring the same region of the sky at
different frequencies and angular scales.
One could expect the amplitude to be obtained
from measurements of temperature anisotropies
on experiments like SuZie (Holzapfel et al. 1997), Ryle
(Saunders 1997), VSA (Jones 1996) or the
upcomming MAP and Planck satellite missions. This will set
strong constraints on the average properties
of the cluster population.

We showed that a relation between the temperature anisotropy
and the mean Comptonization parameter
exists on the average, independent of the cluster
model and IC gas evolution. The influence $p_o$ shown in 
Figure 1b and 3a cancels out in the ratio $\eta$. In a first approximation,
this ratio only varies with the number density
of clusters contributing to SZ effect.
This conclusion relies on the assumption that the cluster population
as a whole is well described by the scaling relations of
eq.~([\ref{Temp}]-[\ref{rc}]). This hypothesis, known
as Weak Self-Similarity Principle (Bower 1997) do not
imply these relations should be accurate for single cluster. For example,
asphericity is important when computing the effect
of a single cluster on the CMB, but it should average
out when considering the effect of the whole cluster population.
But, if electron clumpiness or temperature gradients are shown to be common
in clusters, this would limit the validity of our results.
We do not know how to quantify these effects  at
the present and for this reason we did not include them into our analysis.
Also, let us remark that the influence of different cosmological models 
through $dV/dz$ has not been addressed in this paper.

To conclude, if our cluster model
is applicable to the average cluster population and the number
density of SZ clusters is known, then data on temperature anisotropies can
be used to determine the distortion of the CMB.
On the other hand two independent measurements
of temperature anisotropy and distortion would 
provide an estimate of the number density of Sunyaev--Zel'dovich 
clusters. Comparison with the number
density of X-ray clusters will help to understand
cluster formation and evolution.

{\bf Acknowledgments.}
We thank the referee for his careful reading of the manuscript
that helped to improve the paper substantially.
This research was supported by Spanish German Integrated Actions
HA 97/39 . FAB would like to acknowledge the support of the
Junta de Castilla y Le\'on, grant SA40/97.

\newpage

\newpage

\newpage

\end{document}